\documentclass[final]{aipproc}

\layoutstyle{6x9}

\newcommand\aj{AJ}%
\newcommand\apj{ApJ}%
\newcommand\pra{Phys.~Rev.~A}%

\begin{document}
\title[DZ white dwarfs]{The cool end of the DZ sequence in the SDSS}

\classification{97, 97.10Ex, 97.20.Rp}
\keywords      {Cool white dwarfs, DZ}

\author{D. Koester}{
  address={Institut f\"ur Theoretische Physik und Astrophysik,Universit\"at Kiel, Germany}
}
\author{J. Girven}{
  address={Department of Physics, University of Warwick, Coventry, UK}
}
\author{B. Gaensicke}{
  address={Department of Physics, University of Warwick, Coventry, UK}
}
\author{P. Dufour}{
  address={D\'epartement de Physique, Universit\'e de Montr\'eal, Montr\'eal, QC H3C 3J7, Canada}
}

\begin{abstract}
We report the discovery of cool DZ white dwarfs, which lie in the SDSS
(u-g) vs. (g-r) two-color diagram across and below the main
sequence. These stars represent the extension of the well-known DZ
sequence towards cooler temperatures.
\end{abstract}

\maketitle


\section{Introduction}
Cool He-rich metal polluted white dwarfs (spectral class DZ) in the
SDSS data have been analyzed by \cite{Dufour.Bergeron.ea07}. In the
(u-g) vs. (g-r) diagram (Fig. 1) all but one fall in the region above
the main sequence of normal stars (grey dots). Of the 147 objects
discovered in the SDSS only two have T$_{eff}$ below 6600~K (at 6090
and 4660~K). This suggests that a large number of cooler DZ remains to
be found. We have therefore visually checked all spectra in the region
of QSO colors with borders described by the dashed lines and
identified 17 as cool DZ. The location in the two-color diagram
indicates that the DZ sequence might ``tunnel'' through the main
sequence. Using a new search routine in the region bounded by the
continuous lines, which checks for the characteristic features (most
prominently MgI lines near 5170~\AA), we found another 6 objects, 2 of
which were already in the Dufour sample.

\begin{figure}
  \includegraphics[width=0.65\textwidth]{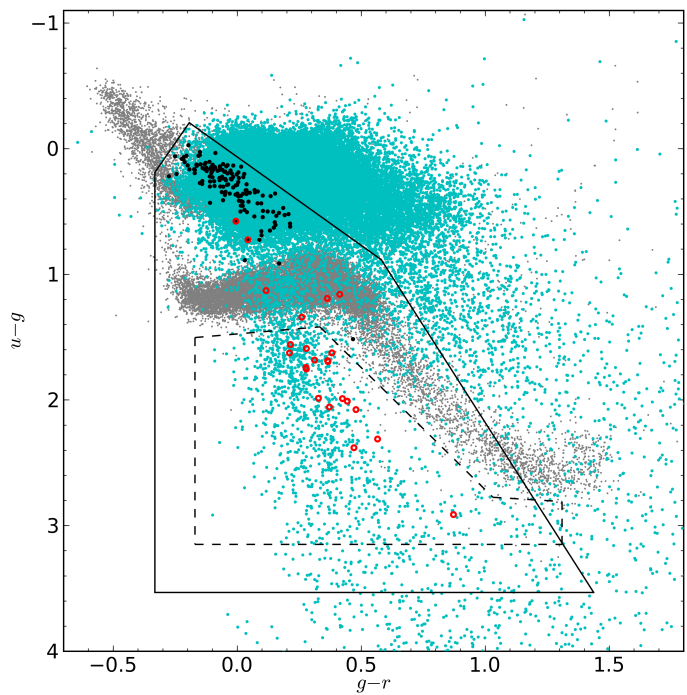} 

\caption{ SDSS Two-color diagram for objects with spectra identified
  as normal stars (grey S-shaped region in the middle of the
  figure), QSO (dots below main sequence), DZ white dwarfs analyzed by
  \cite{Dufour.Bergeron.ea07} (heavy black dots) and cool ``extreme'' DZ
  discovered by our search (small circles).}
\end{figure}

Fig. 2 shows a selection of typical spectra. SDSS1535+1247
(= NLTT40607 = G137-24) was already identified as a DZ and analyzed by
\cite{Kawka.Vennes.ea04} and \cite{Kawka.Vennes06}. It looks extremely
similar to G165-7, shown by \cite{Dufour.Bergeron.ea06}
to be weakly magnetic. That star is also in our sample
(SDSS1330+3029). SDSS0916+2540 has the strongest features of any DZ
known, whereas the remaining two show little more than the strong
asymmetric feature between 5000 and 5200~\AA, which is characteristic
for this group. SDSS1336+3547 is magnetic, with multiple line cores in
many metal lines.

\begin{figure}
  \includegraphics[width=0.60\textwidth]{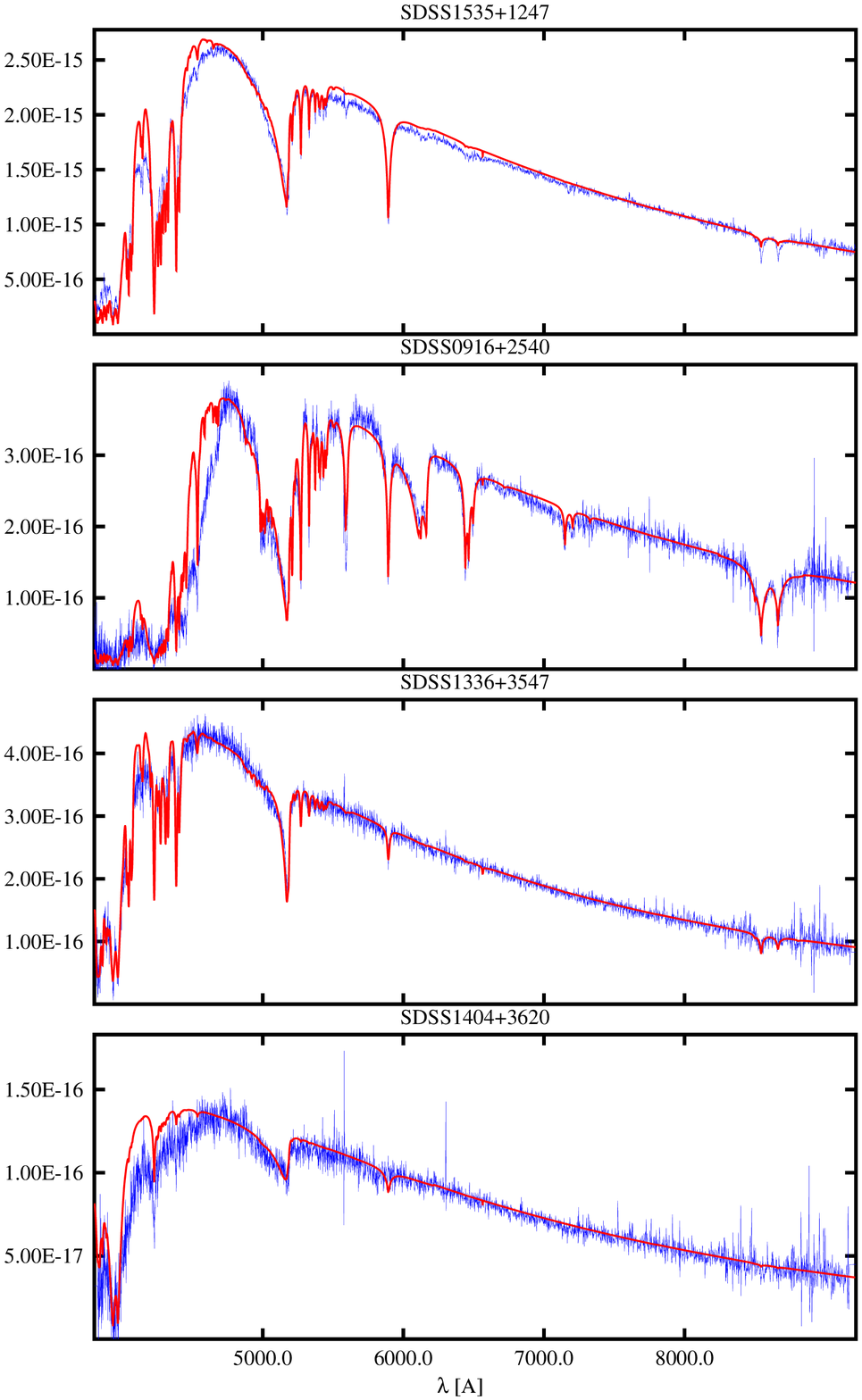} 
\caption{Four stars from our sample of ``extreme'' DZ stars (blue), together
  with theoretical models (red), which at least qualitatively reproduce the
  line features and the continuum slope. If the models are not visible
in black and white print, they fall exactly on the observed spectrum.}
\end{figure}

Our list of extreme DZ is given here

{ \small
\begin{verbatim}
SDSS 0157+0033  SDSS 0205+2155  SDSS 0916+2540  SDSS 0925+3130   
SDSS 0937+5228  SDSS 0956+5912  SDSS 1033+1809  SDSS 1038-0036   
SDSS 1040+2407  SDSS 1043+3516  SDSS 1103+4144  SDSS 1144+1218 
SDSS 1152+1605  SDSS 1234+5208  SDSS 1330+3029  SDSS 1336+3547   
SDSS 1404+3620  SDSS 1421+1843  SDSS 1430-0151  SDSS 1524+4049   
SDSS 1535+1247  SDSS 1546+3009  SDSS 1616+3303
\end{verbatim}
}

\section{Analysis}
All spectral features can be identified with lines from Ca, Mg, Na,
Fe, Ti, and Cr. The broadest lines show strongly asymmetric profiles, which
were in the case of MgI 5169/5174/5185 correctly identified as due to
quasistatic broadening by \cite{Kawka.Vennes.ea04}. The width of these
lines, as well as that of the CaI and CaII resonance lines is far
beyond the range of validity of the impact approximation. We have used
the simple and elegant method of \cite{Walkup.Stewart.ea84}, who
present numerical calculations for the transition range between impact
and quasistatic regime, which can in both limits be easily extended
with the asymptotic formulae. These profiles are reasonable
approximations in many cases. Nevertheless, the current analysis is
only a first attempt and the results are a preliminary estimate of the
general parameters of these objects. Among the problems which will
need to be resolved are

\begin{itemize}
\item The CaI and CaII resonance lines between 3900 and 4250 A are so
  strong that they completely cut off all the flux below ~4500 A. Our
  profiles assume a van der Waals $r^{-6}$ dependence for the
  perturbation energy, which has been shown by
  \cite{Czuchaj.Rebentrost.ea91} to be a very poor approximation for
  the Ca-He interaction. This assumption has to be replaced by a more
  sophisticated potential for each individual line. The current
  profiles are not a satisfactory approximation.

\item The u-g colors suggest that the UV flux near 3500~\AA\ is almost
  totally suppressed in all objects. We have included all expected
  strong lines in the UV -- in particular from FeI and MgI/MgII -- in
  our modeling, which results in a strong blanketing effect in the
  optical. The models qualitatively reproduce the u-g colors, and
  therefore also the blanketing effect. However, the MgI/II resoncance
  lines are expected to be extremely strong, and similar caution
  applies to their profiles as for the CaI/II resonance lines.

\item The upper limit to the H/He ratio is $\approx 10^{-4}$ from the
  non-visibility of H$\alpha$, but the exact amount is unknown. Since
  the broadening of Lyman $\alpha$ by helium leads to very strong
  absorption even into the optical range, this is a further unknown
  free parameter.

\item Finally, the surface gravity can also not be determined from the
  available observations, and we have assumed $\log g = 8$.
\end{itemize}

\section{Results}
For a preliminary analysis we calculated a grid of helium-rich
atmosphere models with T$_{eff}$ between 7000 and 5000~K, log g fixed at
8.0, and variable metal abundances of 13 most commonly detected
elements between Na and Fe, with the ratios kept at approximately
solar values. The hydrogen abundance was fixed at log H/He = -4. A
temperature was determined from the slope of the spectrum, which is in
most cases confirmed by the SDSS magnitudes. In some cases
additionally the ionization balance of CaI/CaII was taken into
account.  The objects in the two-color diagram below the main sequence
are essentially the continuation of the DZ temperature sequence down
to $\approx$ 5400~K.

For the most metal-rich objects individual abundances were adjusted
until a reasonable fit to the spectral features was achieved. As an
example we note  results for two objects with the strongest features\\

\noindent {\bf SDSS 1535+1247:} T$_{eff}$ = 5700 K, [Mg/He] = -7.50,
          [Na/He] = -8.90, [Ca/He] = -8.90, [Fe/He] = -7.60

\noindent {\bf SDSS 0916+2540:} T$_{eff}$ = 5500 K, [Mg/He] = -6.90,
          [Na/He] = -9.25, [Ca/He] = -7.50, [Fe/He] = -7.10.\\

It is obvious that -- while the Mg/Fe ratios are quite similar -- the
relative Ca and Na abundances vary significantly. Since the diffusion
time scales of these elements agree within about 30\%, it is quite
likely that these differences have to be attributed to different
compositions of the accreted material. While the Ca abundance of
SDSS~0916+2540 looks exceptionally high compared to the
Dufour et al. results, the other objects follow the
general trend of their Fig.~9.

The total mass of observed metals in the convection zone of
SDSS~0916+2540 is $\approx 5.2~10^{21}$ g. Adding C, O, Si with the
relative abundances as observed in GD40 \citep{Klein.Jura.ea10} the
total mass of metals is $\approx 1.4~10^{23}$ g. If accretion really
has reached diffusion equilibrium (the diffusion time scales are
typically 300000 yrs) the total accreted mass is of the order of the
Pallas or Vesta masses.

\end{document}